\documentclass[twocolumn,showpacs,preprintnumbers,amsmath,amssymb]{revtex4}

\usepackage{graphicx}
\usepackage{dcolumn}
\usepackage{bm}
\usepackage[tight]{subfigure}

\newcommand{\half}{\textstyle \frac{1}{2}}

\begin{document}
\title{Quantum tunneling induced Kondo effect in single molecular magnets}
\author{C. Romeike}
\author{M. R. Wegewijs}
\author{W. Hofstetter}
\author{H. Schoeller}
\affiliation{
Institut f\"ur Theoretische Physik A, RWTH Aachen, 52056 Aachen,  Germany }

\begin{abstract}
We consider transport through a single-molecule magnet
strongly coupled to metallic electrodes. 
We demonstrate that for half-integer spin of the molecule 
electron- and spin-tunneling \emph{cooperate}
to produce both quantum tunneling of the magnetic moment
and a Kondo effect in the linear conductance. 
The Kondo temperature depends sensitively on the ratio of
the transverse and easy-axis anisotropies in a non-monotonic way.
The magnetic symmetry of the transverse anisotropy 
imposes a selection rule on the total spin for the
occurrence of the Kondo effect which deviates from 
the usual even-odd alternation. 
\end{abstract}

 \pacs{
   72.10.Fk,    
   75.10.Jm,   
   75.30.Gw, 
   75.60.Jk   
}

\maketitle

{\em Introduction.}
Single-molecule magnets (SMMs) such as Mn$_{12}$ or Fe$_{8}$ have
been the focus of intense experimental and theoretical
investigation~\cite{gattorev03}. 
These molecules are characterized by a large spin ($S>1/2$),
easy-axis and transverse anisotropies, and weak intermolecular interaction. 
Molecular-crystal properties are due to an ensemble of single
molecules and exhibit quantum tunneling of
magnetization (QTM) on a mesoscopic scale. 
Recently,  a \emph{single} molecule magnet 
(Mn$_{12}$) was trapped in a nanogap~\cite{Heersche,Jo06} and 
fingerprints of the molecular spin were observed in electron 
transport. 
Furthermore, transport fingerprints of QTM were predicted~\cite{Romeike}
when the individual excitations can be resolved by the temperature.
Using easy-axis anisotropy for magnetic device operation was  
also proposed~\cite{Timm05}.
These works focused on the regime where single electrons charge
and discharge the molecule through weak tunneling.\\
In this Letter we investigate linear transport through a
\emph{half-integer spin} SMM deep inside the blockade 
regime~\cite{Kim04} where the charge on the molecule remains fixed. 
A strong tunnel-coupling to the metallic electrodes induces spin
fluctuations and allows the magnetic moment to tunnel. 
This is remarkable, since for an \emph{isolated} SMM with half-integer
$S$ this is forbidden by time-reversal (TR) symmetry. 
At the same time, the resonant spin-scattering allows electrons to
pass through the SMM: 
the Kondo effect for transport~\cite{Glazman88,Ng88} 
results in a zero bias conductance anomaly 
that has been studied experimentally 
in many systems with small spins (e.g.  quantum
dots~\cite{Goldhaber98,Cronenwett98,Simmel99,Schmid00,vdWiel00}
and single molecules~\cite{Park02,Liang02}).
Such an effect is unexpected in SMMs because  
the $S>1/2$ {\em underscreened} Kondo effect is suppressed by the 
easy-axis anisotropy barrier which freezes the spin along the easy axis. 
However, we find that even a weak transverse anisotropy \emph{induces} a 
pseudo-spin-1/2 Kondo effect. The corresponding Kondo temperature  
is experimentally accessible due to a compensation by  
the large value of the physical spin $S$.  
We perform a scaling analysis~\cite{Anderson70a} for the effective 
pseudo-spin-1/2 model and verify the results by a 
non-perturbative numerical renormalization group (NRG) 
calculation~\cite{Wilson75,Hofstetter00} for the full large-spin 
Hamiltonian. 

\begin{figure}
\includegraphics[scale=0.23]{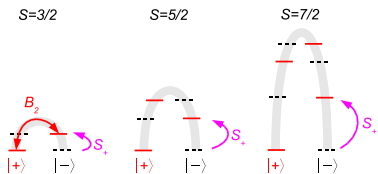}
\caption{\label{fig:levels2} 
Coupling scheme of the ground states including QTM $B_2$ for $S=3/2, 5/2, 7/2$. 
Black (dashed) and red (full) lines denote $|S_z \rangle$ states belonging to the different TR-invariant subspaces. 
}
\end{figure}

{\em Model.}
We consider SMMs which can be described by the following minimal model 
in the limit of strong tunnel-coupling to electron reservoirs 
$H = H_{\text{M}} + H_{\text{K}}$:
\begin{eqnarray}
  \label{eq:ham_mol}
  H_{\text{M}} &=& -D S_{z}^{2}
  - \frac{1}{2} \sum_{n=1}^3 B_{2n}\left ((S^2_{+})^{n}+(S^2_{-})^{n} \right) \\
  \label{eq:ham_kondo}
  H_{\text{K}} &=&  J  \bm{S} \cdot \bm{s} +\sum_{k \sigma} \epsilon_{k \sigma} a_{k \sigma}^{\dag} a_{k \sigma}
 \end{eqnarray}
Here $S_z$ is the projection of the molecule's spin on the easy axis, 
chosen as $z$-axis, $S_\pm = S_x \pm i S_y$,  
and 
$\bm{s} = \sum_{k k'} \sum_{\sigma \sigma'}
  a_{k\sigma}^{\dag} (\frac{1}{2}\bm{\tau})_{\sigma \sigma'}  a_{k'
    \sigma'}$
is the local electron spin in the reservoir 
($\bm{\tau}$ is the Pauli-matrix vector).
As usual, the electronic states $| k \sigma \rangle$ are a linear combination of states of
both physical electrodes~\cite{Glazman88,Ng88}. 
The first term in Eq.~(\ref{eq:ham_mol})
describes the easy-axis magnetic anisotropy of the molecule, 
i.e. the states $|S_z \rangle,S_z=-S,\ldots,S$ are its eigenstates. 
The second term in Eq.~(\ref{eq:ham_mol}) describes transverse
anisotropy perturbations which in general reduce the symmetry to that
of a discrete group of symmetry operations caused by the geometrical
structure of the molecule and its ligands via spin-orbit effects. 
The individual transverse terms written in our model are invariant
under $2n$-fold rotation ($n=1,2,3$) about the easy axis.  
To keep the notation systematic we deviate from
the conventional notation for the anisotropies $E=B_2$ and $C=B_4$. 
Note that the relative strength of the perturbations is $B_{2n} S^{2(n-1)} /D$.
The first term in Eq.~(\ref{eq:ham_kondo}) describes the exchange coupling of the molecular 
spin to the effective reservoir [the second term in Eq.~(\ref{eq:ham_kondo})] and 
thereby transfers charge. 
The coupling is antiferromagnetic, $J>0$, in the blockade regime 
as may be readily shown from the Schrieffer-Wolff
transformation~\cite{SW66}. 
\\
We will show that in a half-integer spin SMM 
the Kondo effect lifts the blockade of both spin-tunneling (due to
TR symmetry) and electron-tunneling (due to energy and charge
quantization). 
Concerning the former effect, 
for half-integer $S>1/2$ the eigenstates of the molecular
Hamiltonian $H_{\text{M}}$ are at least two-fold degenerate
(Kramers doublets)
and are linear combinations of states from only one of the disjoint sets
 $\lbrace | \mp S \pm (2n)k \rangle \rbrace_{k=0,1,2,...}$, 
see Fig.~\ref{fig:levels2}. 
Hence the transverse perturbations $B_{2n}$ can not connect the opposite magnetic
basis states $|\pm S\rangle$ as they do for integer spin
$S$, i.e. QTM is blocked~\cite{Chiolero98}.
However, a Kondo spin-flip process,
Eq.~(\ref{eq:ham_kondo}), can change $S_z$ by {\em one} and connect
the disjoint sets of molecular states.
Thus in cooperation with the QTM terms the molecular spin can be completely
reversed.
This is similar to the co-tunneling of nuclear and electronic spins~\cite{Giraud01}.
Note that in~\cite{Ujsaghy96} a Kondo
effect due to a positive easy-axis anisotropy ($D<0$) was studied  
and no transverse anisotropies were considered.
\\
In the following  we compare situations where either low-
or high-symmetry QTM perturbations dominate. 
This can be achieved experimentally by a  chemical 
modification of the ligands~\cite{Barco05} or 
in transport experiments by changing the binding of the 
molecule to the electrodes, which can be controlled mechanically
in some setups~\cite{Reichert02}.

\begin{figure}[]
\includegraphics[scale =0.19,angle=-0]{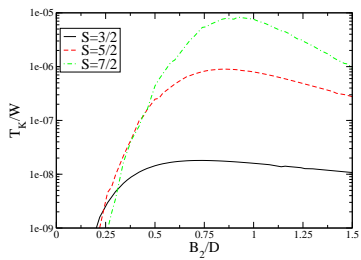}
\caption{\label{fig:tk_nrg_b2} 
Kondo temperature, deduced from the NRG level flow with $J=0.1 W$ and
$D=5 \times 10^{-3} W$, as function of QTM $B_2 $ for $S=3/2,5/2,7/2$. 
}
\end{figure}

\begin{figure}[] 
  \subfigure[][]{\includegraphics[scale=0.23]{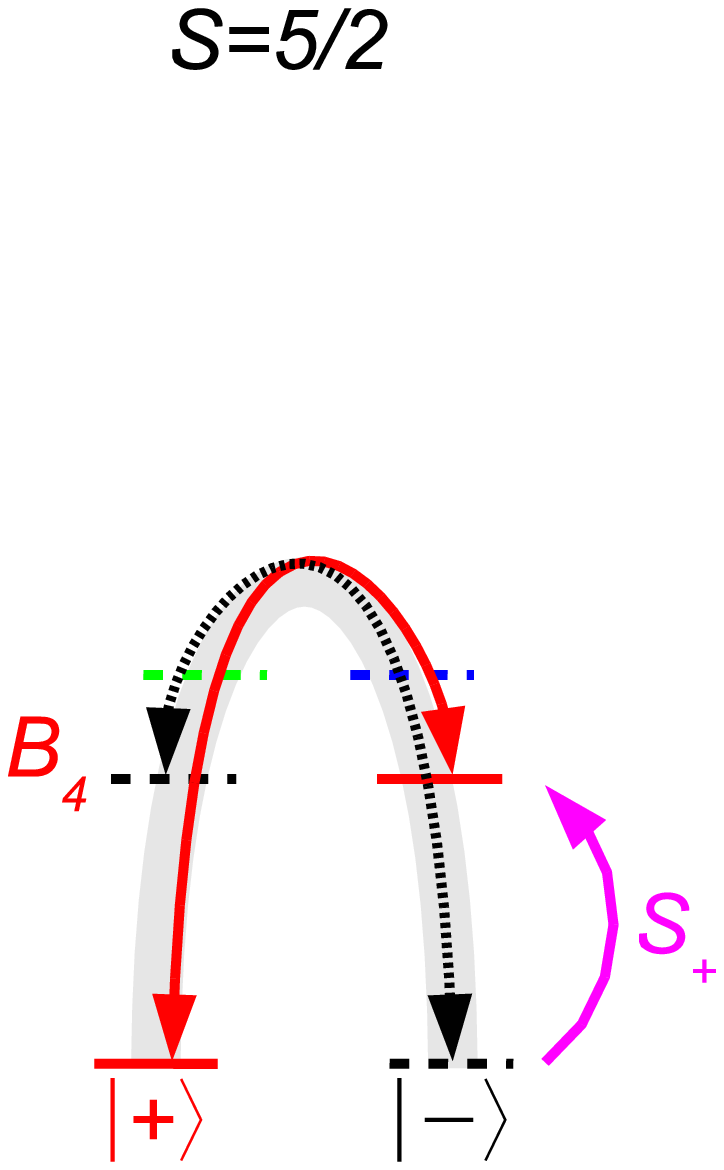}} 
  \subfigure[][]{\includegraphics[scale=0.23]{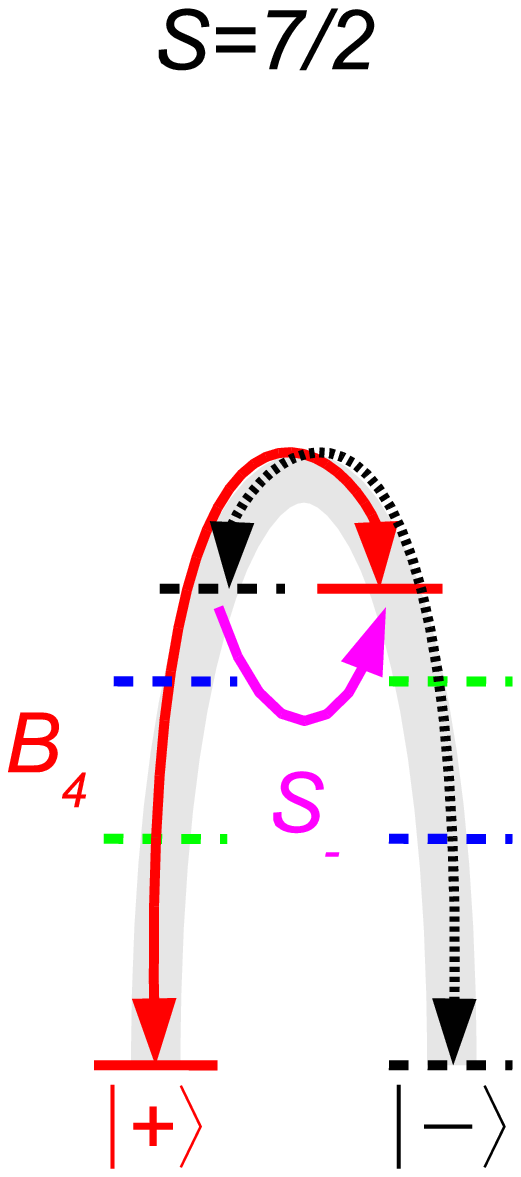}} 
  \subfigure[][]{\includegraphics[scale=0.23]{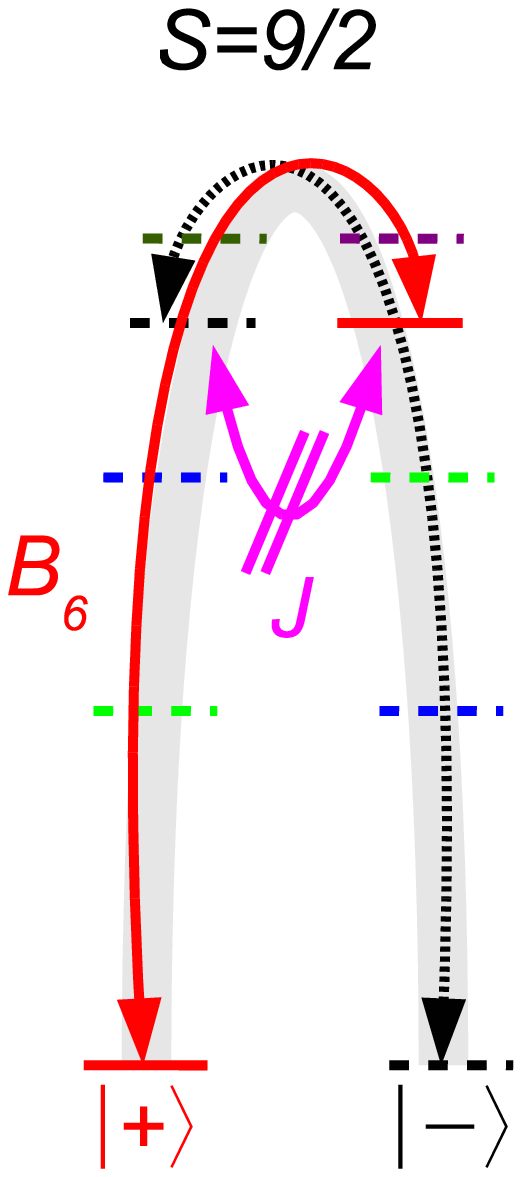}} 
\caption{\label{fig:levels4}  
Scheme for the spin selection rule. 
(a) It is fulfilled for QTM $B_4$ and $S=5/2$ leading to a Kondo 
effect.   
(b) For QTM $B_4$ and $S=7/2$ the product of the effective couplings 
${j_x}{j_y}<0$ and the Kondo effect is suppressed.  
(c) For QTM $B_6$ and spin $S=9/2$. The Kondo  
coupling can not couple the ground state doublet.} 
\end{figure}

\begin{figure}[b] 
\includegraphics[scale =0.19]{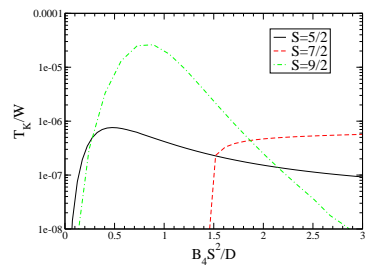} 
\caption{\label{fig:tk_nrg_b4}  
Kondo temperature, deduced from the NRG level flow,  
as function of QTM $B_4$ for $S=5/2,7/2,9/2$. All other parameters as 
in Fig.~\ref{fig:tk_nrg_b2}. 
}
\end{figure}

{\em Poor-man scaling analysis.}
To describe the low energy properties of the above model 
we perform a poor-man scaling analysis~\cite{Anderson70a}. 
This approach leads to similar results as the full NRG
calculations (shown in Fig.~\ref{fig:tk_nrg_b2}, \ref{fig:tk_nrg_b4} and discussed at
the end), but  allows for a detailed discussion of the processes leading 
to Kondo physics. 
We truncate the spectrum to the
twofold degenerate ground state $|\pm \rangle$ and
obtain an effective spin-1/2 Kondo model 
\begin{eqnarray}
  \label{eq:ham_eff}
H_{\text{eff}}=J \sum_{\nu \nu'= \pm} | \nu \rangle \langle \nu' |
\langle \nu | \bm{S} |  \nu'\rangle  \bm{s} = \sum_{i=x,y,z} j_i P_i s_i,
\end{eqnarray}
with the pseudo-spin operators  
$P_\pm = P_x \pm i P_y= |\pm \rangle \langle \mp |$ 
and 
$P_z=(|+\rangle \langle+ | - |-\rangle \langle- |)/2$.
The effective exchange constants depend on $ B_{2n}$, $D$ and $S$
through 
\begin{eqnarray}
  \label{eq:jz}
  j_z     &=&  2 J \langle +   | S_z | + \rangle\,>0  \\
  \label{eq:jxy}
  j_{x,y} &=&  J \langle +   | S_{+} \pm S_{-} | - \rangle 
\end{eqnarray}
which we have calculated numerically.
Importantly, these constants are \emph{completely} anisotropic, except
for special cases.
The scaling equations are 
($2W=$ conduction electron bandwidth, $\rho=$ density of states):
\begin{eqnarray}
  \label{eq:sca}
  \frac{d {j_\alpha}}{d \ln W} = - \rho {j_\beta} {j_\gamma}
\end{eqnarray}
where $\alpha,\beta,\gamma$ are cyclic permutations of $x,y,z$~\cite{Zawadowski80}.
Specification of any two scaling invariants
${j_\alpha}^2 - {j_\beta}^2,\alpha \ne \beta$, defines a 3-dimensional scaling curve.
Inversion of any pair of $j_\alpha,j_\beta$ leaves the scaling equations invariant,
whereas inverting a single one reverses the flow.
Interestingly, all scaling trajectories flow to the strong coupling
limit except those in planes of uni-axial symmetry, 
$|j_\alpha| = |j_\beta| < |j_\gamma|$ with $j_\alpha j_\beta j_\gamma < 0$.
In the latter case one has a ferromagnetic fixed line which 
is unstable with respect to infinitesimal 
perturbations perpendicular to it which are typically present in our
model. 
If the effective exchange constants lie close to
this line the Kondo temperature will thus be strongly suppressed.
If the Kondo effect occurs and $|j_z| \ge |j_x| \ge |j_y|$ 
with $|j_z| \ne |j_y| $, we find 
for the Kondo temperature
(defined here as the scale where the first coupling constant diverges)
\begin{eqnarray}
  \label{eq:tk}
    \ln (T_K / W_\text{eff}) =
- \frac {
  \text{cs}^{- 1} 
  \left(
    \frac{|j_y|}{\sqrt{{j_z}^2 - {j_y}^2}}
    \Big\lvert
    \frac{{j_z}^2 - {j_x}^2}{{j_z}^2 - {j_y}^2} 
   \right)}
  {\rho \sqrt{{j_z}^2 - {j_y}^2}}. 
\end{eqnarray}
Here $\text{cs}^{-1}(u|m)$ is the inverse of the elliptic integral
$\text{cs}(u|m)$, see~\cite{Abramowitz}.
In the uni-axial planes $|j_z|=|j_x|$ or $|j_x|=|j_y|$ Eq.~(\ref{eq:tk})
reduces to the well-known expressions for easy-axis
anisotropy~\cite{Anderson70b,Tsvelick83} since 
$\textrm{cs}^{-1}(u|0)= \arctan(1/u)$
and
$\textrm{cs}^{-1}(u|1)= \textrm{arctanh} (\sqrt{1+u^2})$.
In this model the upper bound  of the Kondo scale $T_K$ is the 
energy separation between Kramers-degenerate ground and first excited
state of the isolated molecule $W_\text{eff}(D,\{B_{2n}\})$ which is 
of the order $0.1~\text{meV} \sim 1~\text{K}$.\\
According to Eq.~(\ref{eq:jxy}), the exchange couplings
$j_{x,y}$ are generated by spin-tunneling. This
gives $j_z > |j_x|,|j_y|$ for not
too strong QTM. Thus, the only case where the Kondo
effect can not be observed is $|j_x|=|j_y|$ and $j_x j_y<0$, which
using Eq.~(\ref{eq:jxy}) gives 
$\langle +|S_+|-\rangle = 0$. This means that the Kondo effect is not 
observable when the spin raising operator of the original molecular spin 
can not flip the pseudo-spin from the down to the up value, an intuitively
quite obvious condition. Most importantly, as we will illustrate in the
following, this condition can be checked very easily provided the spin
and the symmetry of the molecular magnet are given. If a $B_{2n}$ 
quantum tunneling term is present, we get
\begin{eqnarray}
 \label{eq:cond}
 \langle +|S_+|-\rangle \ne 0\quad \Leftrightarrow \quad
\frac {2S-1}{2n}= \text{Integer}
\end{eqnarray}
as a condition for the observability of the Kondo effect in molecular
magnets with weak QTM.\\
We first consider the limit of a dominant low-symmetry QTM term, $B_2
\gg B_4 S^2, B_6 S^4$.
We \emph{always} find a spin-1/2 Kondo effect, see
Fig.~\ref{fig:tk_nrg_b2}, because
the three couplings are different except for $B_2=D$. 
At that point Eq. (\ref{eq:ham_mol}) can be rewritten as
$H_{\text{M}}=2D S_y^2 + \textrm{const}$.
The resulting uni-axial symmetric couplings $|j_z|=|j_x| > |j_y|$
allow for a flow to the strong coupling fixed point, also in this case. 
The Kondo temperature, shown in Fig.~\ref{fig:tk_nrg_b2},
has a non-monotonic dependence on $B_2/D$, which is enhanced with
increasing $S$.
For $B_2 \ll D $ (weak QTM), the criterion
(\ref{eq:cond}) applies and is always fulfilled for any half-integer spin. 
The states forming the two ground states $|\pm\rangle$
are connected by the spin raising operator by $S-1/2$  QTM processes 
(each contributing a factor $\propto B_2/D$) and one co-tunneling process,
see Fig.~\ref{fig:levels2}.
Therefore, the $B_2/D$ dependence, estimated
from Eqs.~(\ref{eq:jz},\ref{eq:jxy}), is
$|j_z| \approx 2SJ \gg |j_x| \approx |j_y| \propto
(B_2/D)^{S-\frac{1}{2}}$.
Using (\ref{eq:tk}) this gives for the Kondo temperature
\begin{eqnarray}
  T_K^{B_2 \ll D} / W_\text{eff}
  \propto  e^{-\frac{1}{2\rho J}(1-\frac{1}{2S})\ln{\frac{D}{B_2}}}.
  \label{eq:tk_weak}
\end{eqnarray}
The exponent becomes $S$-independent for $S \gg 1$.
However, the complicated spin-dependent prefactor left out
in Eq.~(\ref{eq:tk_weak}) decreases with $S$ stronger than
$W_{\text{eff}}$ increases:
in this regime the Kondo temperature therefore decreases with increasing spin. 
Interestingly, this tendency changes for larger quantum tunneling.
Near $B_2= D$ the perpendicular couplings
dominate and grow with increasing $S$: 
$j_{x,z}= J \sqrt{S(S+1)+1/4}$ and $j_y = J$ for $B_2 = D$.
From Eq. (\ref{eq:tk}) one obtains by expanding in $j_y/j_z \propto 1/S$
 an enhancement of $T_K$ with $S \gg 1$:
\begin{eqnarray}
 T_K^{B_2=D} /  W_\text{eff}
  \approx e^{ -\frac{\pi}{2} \frac{1}{\rho J \sqrt{S(S+1)+\frac{1}{4}}} }.
\end{eqnarray}
In this expression there are two competing factors:
increasing $B_2/D$ enhances  
the r.h.s. of Eq.~(\ref{eq:tk}) which is maximal at $B_{2}=D$, 
but simultaneously reduces the splitting $W_{\text{eff}}$
between ground and excited state from $(2S-1)D$ to $4D$. 
Hence the maximal Kondo temperature occurs for a value $B_2/D <1$.
Finally, for $B_2 \gg D$ the Kondo temperature is suppressed again
since one coupling becomes dominant: 
$|j_x| \approx J \gg |j_y|,|j_z|$. We note that in the case 
$B_2=D$ the molecule can not be considered as a molecular magnet 
 and it is only discussed here to explain the tendency of the
increase of the Kondo temperature with increasing spin for strong
quantum tunneling. 
We suggest to study SMMs with moderate 
quantum tunneling, as e.g.  Fe$_8^{\text{(III)}}$ with $D=0.27$~K  
and  $B_2=0.046$~K~\cite{Sangregorio97}.
For the above mechanism to be relevant the
 spin $S=10$ needs to be changed to an half-integer value by changing
the charge via a gate electrode.
The value of the Kondo exchange coupling depends on the details of the
adjacent charge states, e.g. changes in anisotropies and total spin.
For a quantitative calculation further input  
from experiment and ab-initio calculations (e.g. \cite{Park04add})  
is needed. \\
Now we consider a dominant QTM perturbation of  
higher symmetry, $B_4 \gg B_2 / S^2, B_6 S^2$.  
For $B_2=B_6=0$ we have 4 disjoint subsets of basis states 
which cannot be connected by the QTM term, e.g.   
for $B_4 S^2 \ll D$ the ground states are linear 
combinations of $\lbrace | \mp S \pm 4k \rangle
\rbrace_{k=0,1,2,...}$, see Fig.~\ref{fig:levels4}(a,b).  
While condition (\ref{eq:cond}) is fulfilled  
for $S=5/2 + 2m$, ($m=0,1,\dots$), Fig.~\ref{fig:levels4}(a),  
it is violated for spin $S=7/2 + 2m$.  In the latter case, only  
the spin lowering operator can increase the pseudo-spin, Fig.~\ref{fig:levels4}(b).  
However, with increasing $B_4 S^2/D$ a level crossing between the ground and 
excited state of different symmetry results in a sharp change in  
the Kondo temperature. Hence $B_4$ induces a 
quantum-phase transition, see Fig.~\ref{fig:tk_nrg_b4} for $S=7/2$, c.f. ~\cite{Romeike}.  
\\
Finally, we mention the possibility that the Kondo coupling can not 
connect two TR-invariant subsets which leads to a complete
vanishing of the effective coupling constants ${j_{x,y}}=0$. 
The lowest order QTM where this effect takes place is of 
sixth order: in general the Kondo coupling cannot overcome 
the ``mismatch'' $|\Delta S_z| = 6$, see Fig.~\ref{fig:levels4}(c). 
In this case only for intermediate or strong $B_6$ a 
level crossing may give rise to
ground states which support a Kondo effect.

{\em Numerical RG.}
We use Wilsons's NRG~\cite{Wilson75,Hofstetter00} to check the results
of our scaling analysis taking into account the full model (i.e. no
truncation to a ground state doublet is made). 
As input parameters we take  
$J=0.1 W$, 
number of states $N=1500$, 
discretization $\Lambda=2$ and 
$D= 5 \times 10^{-3} W$. 
Since the original procedure has been formulated for a
spin-1/2 Kondo model we have modified it to incorporate
an arbitrary spin of the impurity.
We analyzed the RG level flow as function of iteration number $N_\text{iter}$ in
order to determine the low-temperature fixed point and the Kondo 
temperature which is defined as the energy scale where the 
crossover to strong coupling takes place. 
For half-integer spin $S$ and $D>0$
we observe a flow to the strong coupling fixed point only for
a QTM perturbation $B_{2n} \ne 0$, as expected from the above scaling analysis. 
The Kondo temperature for spin $S=3/2, 5/2,
7/2$ and dominant $B_2$ QTM
 is plotted in Fig.~\ref{fig:tk_nrg_b2}. 
It shows in good qualitative agreement with the scaling results the 
discussed non-monotonic behavior as function of the QTM.
For dominant $B_4$ QTM
the Kondo temperature is plotted for spin $S=5/2, 7/2, 9/2$
in Fig.~\ref{fig:tk_nrg_b4}.
The mapping onto a pseudo-spin-1/2 system is valid as long as there is no
crossing of levels 
and $T_K$ does not exceed the gap to the first excited state.
The former is the case for the experimentally most
 relevant regime $B_{2n} S^{2(n-1)} < D$. 

{\em Discussion.}
Consequently the observation of Kondo-tunneling through SMMs
requires a judicious selection of three quantities: 
(i) the  total spin must be half-integer; 
(ii) the dominant QTM perturbation should be moderate 
$B_{2n} S^{2n-2} \lesssim D$ and
(iii) the total spin should satisfy the selection rule (\ref{eq:cond})
if a high-symmetry QTM term dominates.
In experiments where the charge state, and hence the spin state, can
be controlled by a gate voltage, the Kondo effect can only be seen
in every $(2n)$ subsequent Coulomb diamond  for molecules
with a dominant $(2n)$ QTM perturbation.
This assumes that the spin $S$ increases/decreases by $\half$ every
next charge state and has to be contrasted to the even/odd alternation of the
Kondo effect usually observed in quantum dots.
It would be of interest to mechanically alter the symmetry of the 
magnetic core of the SMM \emph{in situ}, e.g. in a mechanically
controlled break-junction setup~\cite{Reichert02}, and thereby 
suppress/enhance the Kondo effect. 

{\em Summary.} 
Using scaling and numerical renormalization group techniques 
we have found that  
spin- and electron-tunneling become correlated 
in half-integer spin magnetic molecules which are strongly coupled to 
electrodes. 
The spin-1/2 Kondo anomaly in the linear conductance signals 
the externally induced tunneling of the magnetization of the molecule. 
The Kondo temperature shows a non-monotonic dependence on the 
relative strength of the transverse magnetic anisotropy of the molecule. 
Importantly, the large spin of SMMs is found to compensate for 
anisotropy effects which are expected to suppress Kondo physics. 
The symmetry of the transverse anisotropy imposes a selection rule: 
the Kondo effect only occurs for selected values of 
the molecular spin. 

We would like to thank K. Kikoin, M. Kiselev, J. Kortus, D. Loss, J. Martinek,  
P. Nozi\`eres, H. Park, H. van der Zant and A. Zawadowski for discussions.  
We acknowledge financial support through the EU RTN Spintronics program
HPRN-CT-2002-00302 and the FZ J\"ulich via the virtual institute IFMIT.
\bibliographystyle{apsrev}
\bibliography{Kondo}

\begin{thebibliography}{30}
\expandafter\ifx\csname natexlab\endcsname\relax\def\natexlab#1{#1}\fi
\expandafter\ifx\csname bibnamefont\endcsname\relax
  \def\bibnamefont#1{#1}\fi
\expandafter\ifx\csname bibfnamefont\endcsname\relax
  \def\bibfnamefont#1{#1}\fi
\expandafter\ifx\csname citenamefont\endcsname\relax
  \def\citenamefont#1{#1}\fi
\expandafter\ifx\csname url\endcsname\relax
  \def\url#1{\texttt{#1}}\fi
\expandafter\ifx\csname urlprefix\endcsname\relax\def\urlprefix{URL }\fi
\providecommand{\bibinfo}[2]{#2}
\providecommand{\eprint}[2][]{\url{#2}}

\bibitem[{\citenamefont{Gatteschi and Sessoli}(2003)}]{gattorev03}
\bibinfo{author}{\bibfnamefont{D.}~\bibnamefont{Gatteschi}} \bibnamefont{and}
  \bibinfo{author}{\bibfnamefont{R.}~\bibnamefont{Sessoli}},
  \bibinfo{journal}{Angew. Chem. Int. Ed.} \textbf{\bibinfo{volume}{42}},
  \bibinfo{pages}{268} (\bibinfo{year}{2003}), \bibinfo{note}{and references
  therein}.

\bibitem[{\citenamefont{Heersche et~al.}()\citenamefont{Heersche, de~Groot,
  Folk, van~der Zant, Romeike, Wegewijs, Zobbi, Barreca, Tondello, and
  Cornia}}]{Heersche}
\bibinfo{author}{\bibfnamefont{H.}~\bibnamefont{Heersche}},
  \bibinfo{author}{\bibfnamefont{Z.}~\bibnamefont{de~Groot}},
  \bibinfo{author}{\bibfnamefont{J.~A.} \bibnamefont{Folk}},
  \bibinfo{author}{\bibfnamefont{H.~S.~J.} \bibnamefont{van~der Zant}},
  \bibinfo{author}{\bibfnamefont{C.}~\bibnamefont{Romeike}},
  \bibinfo{author}{\bibfnamefont{M.~R.} \bibnamefont{Wegewijs}},
  \bibinfo{author}{\bibfnamefont{L.}~\bibnamefont{Zobbi}},
  \bibinfo{author}{\bibfnamefont{D.}~\bibnamefont{Barreca}},
  \bibinfo{author}{\bibfnamefont{E.}~\bibnamefont{Tondello}}, \bibnamefont{and}
  \bibinfo{author}{\bibfnamefont{A.}~\bibnamefont{Cornia}},
  \bibinfo{note}{cond-mat/0510732,Phys. Rev. Lett. (to be published)}.

\bibitem[{\citenamefont{Jo et~al.}()\citenamefont{Jo, Grose, Deshmukh,
  Rumberger, Hendrickson, Long, Park, and Ralph}}]{Jo06}
\bibinfo{author}{\bibfnamefont{M.-H.} \bibnamefont{Jo}},
  \bibinfo{author}{\bibfnamefont{J.~E.} \bibnamefont{Grose}},
  \bibinfo{author}{\bibfnamefont{M.~M.} \bibnamefont{Deshmukh}},
  \bibinfo{author}{\bibfnamefont{J.~J. S.~M.} \bibnamefont{Rumberger}},
  \bibinfo{author}{\bibfnamefont{D.~N.} \bibnamefont{Hendrickson}},
  \bibinfo{author}{\bibfnamefont{J.~R.} \bibnamefont{Long}},
  \bibinfo{author}{\bibfnamefont{H.}~\bibnamefont{Park}}, \bibnamefont{and}
  \bibinfo{author}{\bibfnamefont{D.~C.} \bibnamefont{Ralph}},
  \bibinfo{note}{cond-mat/0603276}.

\bibitem[{\citenamefont{Romeike et~al.}()\citenamefont{Romeike, Wegewijs, and
  Schoeller}}]{Romeike}
\bibinfo{author}{\bibfnamefont{C.}~\bibnamefont{Romeike}},
  \bibinfo{author}{\bibfnamefont{M.~R.} \bibnamefont{Wegewijs}},
  \bibnamefont{and}
  \bibinfo{author}{\bibfnamefont{H.}~\bibnamefont{Schoeller}},
  \bibinfo{note}{cond-mat/0511391,Phys. Rev. Lett. (to be published)}.

\bibitem[{\citenamefont{Timm and Elste}()}]{Timm05}
\bibinfo{author}{\bibfnamefont{C.}~\bibnamefont{Timm}} \bibnamefont{and}
  \bibinfo{author}{\bibfnamefont{F.}~\bibnamefont{Elste}},
  \bibinfo{note}{cond-mat/0511291,Phys. Rev. B (to be published)}.

\bibitem[{\citenamefont{Kim and Kim}(2004)}]{Kim04}
\bibinfo{author}{\bibfnamefont{G.-H.} \bibnamefont{Kim}} \bibnamefont{and}
  \bibinfo{author}{\bibfnamefont{T.-S.} \bibnamefont{Kim}},
  \bibinfo{journal}{Phys.\ Rev.\ Lett.} \textbf{\bibinfo{volume}{92}},
  \bibinfo{pages}{137203} (\bibinfo{year}{2004}).

\bibitem[{\citenamefont{Glazman and Raikh}(1988)}]{Glazman88}
\bibinfo{author}{\bibfnamefont{L.~I.} \bibnamefont{Glazman}} \bibnamefont{and}
  \bibinfo{author}{\bibfnamefont{M.~E.} \bibnamefont{Raikh}},
  \bibinfo{journal}{JETP\ Lett.} \textbf{\bibinfo{volume}{47}},
  \bibinfo{pages}{452} (\bibinfo{year}{1988}).

\bibitem[{\citenamefont{Ng and Lee}(1988)}]{Ng88}
\bibinfo{author}{\bibfnamefont{T.~K.} \bibnamefont{Ng}} \bibnamefont{and}
  \bibinfo{author}{\bibfnamefont{P.~A.} \bibnamefont{Lee}},
  \bibinfo{journal}{Phys.\ Rev.\ Lett.} \textbf{\bibinfo{volume}{61}},
  \bibinfo{pages}{1768} (\bibinfo{year}{1988}).

\bibitem[{\citenamefont{Goldhaber-Gordon
  et~al.}(1998)\citenamefont{Goldhaber-Gordon, Shtrikman, Mahalu,
  Abusch-Magder, Meirav, and Kastner}}]{Goldhaber98}
\bibinfo{author}{\bibfnamefont{D.}~\bibnamefont{Goldhaber-Gordon}},
  \bibinfo{author}{\bibfnamefont{H.}~\bibnamefont{Shtrikman}},
  \bibinfo{author}{\bibfnamefont{D.}~\bibnamefont{Mahalu}},
  \bibinfo{author}{\bibfnamefont{D.}~\bibnamefont{Abusch-Magder}},
  \bibinfo{author}{\bibfnamefont{U.}~\bibnamefont{Meirav}}, \bibnamefont{and}
  \bibinfo{author}{\bibfnamefont{M.~A.} \bibnamefont{Kastner}},
  \bibinfo{journal}{Nature (London)} \textbf{\bibinfo{volume}{391}},
  \bibinfo{pages}{156} (\bibinfo{year}{1998}).

\bibitem[{\citenamefont{Cronenwett et~al.}(1998)\citenamefont{Cronenwett,
  Oosterkamp, and Kouwenhoven}}]{Cronenwett98}
\bibinfo{author}{\bibfnamefont{S.~M.} \bibnamefont{Cronenwett}},
  \bibinfo{author}{\bibfnamefont{T.~H.} \bibnamefont{Oosterkamp}},
  \bibnamefont{and} \bibinfo{author}{\bibfnamefont{L.~P.}
  \bibnamefont{Kouwenhoven}}, \bibinfo{journal}{Science}
  \textbf{\bibinfo{volume}{281}}, \bibinfo{pages}{540} (\bibinfo{year}{1998}).

\bibitem[{\citenamefont{Simmel et~al.}(1999)\citenamefont{Simmel, Blick,
  Kotthaus, Wegscheider, and Bichler}}]{Simmel99}
\bibinfo{author}{\bibfnamefont{F.}~\bibnamefont{Simmel}},
  \bibinfo{author}{\bibfnamefont{R.~H.} \bibnamefont{Blick}},
  \bibinfo{author}{\bibfnamefont{J.~P.} \bibnamefont{Kotthaus}},
  \bibinfo{author}{\bibfnamefont{W.}~\bibnamefont{Wegscheider}},
  \bibnamefont{and} \bibinfo{author}{\bibfnamefont{M.}~\bibnamefont{Bichler}},
  \bibinfo{journal}{Phys.\ Rev.\ Lett.} \textbf{\bibinfo{volume}{83}},
  \bibinfo{pages}{804} (\bibinfo{year}{1999}).

\bibitem[{\citenamefont{Schmid et~al.}(2000)\citenamefont{Schmid, Weis, Eberl,
  and v.~Klitzing}}]{Schmid00}
\bibinfo{author}{\bibfnamefont{J.}~\bibnamefont{Schmid}},
  \bibinfo{author}{\bibfnamefont{J.}~\bibnamefont{Weis}},
  \bibinfo{author}{\bibfnamefont{K.}~\bibnamefont{Eberl}}, \bibnamefont{and}
  \bibinfo{author}{\bibfnamefont{K.}~\bibnamefont{v.~Klitzing}},
  \bibinfo{journal}{Phys.\ Rev.\ Lett.} \textbf{\bibinfo{volume}{84}},
  \bibinfo{pages}{5824} (\bibinfo{year}{2000}).

\bibitem[{\citenamefont{van~der Wiel et~al.}(2000)\citenamefont{van~der Wiel,
  de~Franceschi, Fujisawa, Elzerman, Tarucha, and Kouwenhoven}}]{vdWiel00}
\bibinfo{author}{\bibfnamefont{W.~G.} \bibnamefont{van~der Wiel}},
  \bibinfo{author}{\bibfnamefont{S.}~\bibnamefont{de~Franceschi}},
  \bibinfo{author}{\bibfnamefont{T.}~\bibnamefont{Fujisawa}},
  \bibinfo{author}{\bibfnamefont{J.~M.} \bibnamefont{Elzerman}},
  \bibinfo{author}{\bibfnamefont{S.}~\bibnamefont{Tarucha}}, \bibnamefont{and}
  \bibinfo{author}{\bibfnamefont{L.~P.} \bibnamefont{Kouwenhoven}},
  \bibinfo{journal}{Science} \textbf{\bibinfo{volume}{289}},
  \bibinfo{pages}{2105} (\bibinfo{year}{2000}).

\bibitem[{\citenamefont{Park et~al.}(2002)\citenamefont{Park, Pasupathy,
  Goldsmith, Chang, Yaish, Petta, Rinkoski, Sethna, Abru{\~n}a, McEuen
  et~al.}}]{Park02}
\bibinfo{author}{\bibfnamefont{J.}~\bibnamefont{Park}},
  \bibinfo{author}{\bibfnamefont{A.~N.} \bibnamefont{Pasupathy}},
  \bibinfo{author}{\bibfnamefont{J.~I.} \bibnamefont{Goldsmith}},
  \bibinfo{author}{\bibfnamefont{C.}~\bibnamefont{Chang}},
  \bibinfo{author}{\bibfnamefont{Y.}~\bibnamefont{Yaish}},
  \bibinfo{author}{\bibfnamefont{J.~R.} \bibnamefont{Petta}},
  \bibinfo{author}{\bibfnamefont{M.}~\bibnamefont{Rinkoski}},
  \bibinfo{author}{\bibfnamefont{J.~P.} \bibnamefont{Sethna}},
  \bibinfo{author}{\bibfnamefont{H.~D.} \bibnamefont{Abru{\~n}a}},
  \bibinfo{author}{\bibfnamefont{P.~L.} \bibnamefont{McEuen}},
  \bibnamefont{et~al.}, \bibinfo{journal}{Nature}
  \textbf{\bibinfo{volume}{417}}, \bibinfo{pages}{722} (\bibinfo{year}{2002}).

\bibitem[{\citenamefont{Liang et~al.}(2002)\citenamefont{Liang, Shores,
  Bockrath, Long, and Park}}]{Liang02}
\bibinfo{author}{\bibfnamefont{W.}~\bibnamefont{Liang}},
  \bibinfo{author}{\bibfnamefont{M.~P.} \bibnamefont{Shores}},
  \bibinfo{author}{\bibfnamefont{M.}~\bibnamefont{Bockrath}},
  \bibinfo{author}{\bibfnamefont{J.~R.} \bibnamefont{Long}}, \bibnamefont{and}
  \bibinfo{author}{\bibfnamefont{H.}~\bibnamefont{Park}},
  \bibinfo{journal}{Nature} \textbf{\bibinfo{volume}{417}},
  \bibinfo{pages}{725} (\bibinfo{year}{2002}).

\bibitem[{\citenamefont{Anderson}(1970)}]{Anderson70a}
\bibinfo{author}{\bibfnamefont{P.~W.} \bibnamefont{Anderson}},
  \bibinfo{journal}{J. Phys. C} \textbf{\bibinfo{volume}{3}},
  \bibinfo{pages}{2436} (\bibinfo{year}{1970}).

\bibitem[{\citenamefont{Wilson}(1975)}]{Wilson75}
\bibinfo{author}{\bibfnamefont{K.~G.} \bibnamefont{Wilson}},
  \bibinfo{journal}{Rev.\ Mod.\ Phys.} \textbf{\bibinfo{volume}{47}},
  \bibinfo{pages}{773} (\bibinfo{year}{1975}).

\bibitem[{\citenamefont{Hofstetter}(2000)}]{Hofstetter00}
\bibinfo{author}{\bibfnamefont{W.}~\bibnamefont{Hofstetter}},
  \bibinfo{journal}{Phys.\ Rev.\ Lett.} \textbf{\bibinfo{volume}{85}},
  \bibinfo{pages}{1508} (\bibinfo{year}{2000}).

\bibitem[{\citenamefont{Schrieffer and Wolff}(1966)}]{SW66}
\bibinfo{author}{\bibfnamefont{J.~R.} \bibnamefont{Schrieffer}}
  \bibnamefont{and} \bibinfo{author}{\bibfnamefont{P.~A.} \bibnamefont{Wolff}},
  \bibinfo{journal}{Phys.\ Rev.} \textbf{\bibinfo{volume}{149}},
  \bibinfo{pages}{491} (\bibinfo{year}{1966}).

\bibitem[{\citenamefont{Chiolero and Loss}(1998)}]{Chiolero98}
\bibinfo{author}{\bibfnamefont{A.}~\bibnamefont{Chiolero}} \bibnamefont{and}
  \bibinfo{author}{\bibfnamefont{D.}~\bibnamefont{Loss}},
  \bibinfo{journal}{Phys.\ Rev.\ Lett.} \textbf{\bibinfo{volume}{80}},
  \bibinfo{pages}{169} (\bibinfo{year}{1998}).

\bibitem[{\citenamefont{Giraud et~al.}(2001)\citenamefont{Giraud, Wernsdorfer,
  Tkachuk, Mailly, and Barbara}}]{Giraud01}
\bibinfo{author}{\bibfnamefont{R.}~\bibnamefont{Giraud}},
  \bibinfo{author}{\bibfnamefont{W.}~\bibnamefont{Wernsdorfer}},
  \bibinfo{author}{\bibfnamefont{A.~M.} \bibnamefont{Tkachuk}},
  \bibinfo{author}{\bibfnamefont{D.}~\bibnamefont{Mailly}}, \bibnamefont{and}
  \bibinfo{author}{\bibfnamefont{B.}~\bibnamefont{Barbara}},
  \bibinfo{journal}{Phys.\ Rev.\ Lett.} \textbf{\bibinfo{volume}{87}},
  \bibinfo{pages}{057203} (\bibinfo{year}{2001}).

\bibitem[{\citenamefont{\'Ujs\'aghy et~al.}(1996)\citenamefont{\'Ujs\'aghy,
  Zawadowski, and Gyorffy}}]{Ujsaghy96}
\bibinfo{author}{\bibfnamefont{O.}~\bibnamefont{\'Ujs\'aghy}},
  \bibinfo{author}{\bibfnamefont{A.}~\bibnamefont{Zawadowski}},
  \bibnamefont{and} \bibinfo{author}{\bibfnamefont{B.~L.}
  \bibnamefont{Gyorffy}}, \bibinfo{journal}{Phys.\ Rev.\ Lett.}
  \textbf{\bibinfo{volume}{76}}, \bibinfo{pages}{2378} (\bibinfo{year}{1996}).

\bibitem[{\citenamefont{Barco et~al.}(2005)\citenamefont{Barco, Kent, Hill,
  North, Dalal, Rumberger, Hendrickson, Chakov, and Christou}}]{Barco05}
\bibinfo{author}{\bibfnamefont{E.}~\bibnamefont{Barco}},
  \bibinfo{author}{\bibfnamefont{A.~D.} \bibnamefont{Kent}},
  \bibinfo{author}{\bibfnamefont{S.}~\bibnamefont{Hill}},
  \bibinfo{author}{\bibfnamefont{J.~M.} \bibnamefont{North}},
  \bibinfo{author}{\bibfnamefont{N.~S.} \bibnamefont{Dalal}},
  \bibinfo{author}{\bibfnamefont{E.~M.} \bibnamefont{Rumberger}},
  \bibinfo{author}{\bibfnamefont{D.~N.} \bibnamefont{Hendrickson}},
  \bibinfo{author}{\bibfnamefont{N.}~\bibnamefont{Chakov}}, \bibnamefont{and}
  \bibinfo{author}{\bibfnamefont{G.}~\bibnamefont{Christou}},
  \bibinfo{journal}{J. L. Temp. Phys.} \textbf{\bibinfo{volume}{140}},
  \bibinfo{pages}{119} (\bibinfo{year}{2005}).

\bibitem[{\citenamefont{Reichert et~al.}(2002)\citenamefont{Reichert, Ochs,
  Beckmann, Weber, Mayor, and v.~L\"ohneysen}}]{Reichert02}
\bibinfo{author}{\bibfnamefont{J.}~\bibnamefont{Reichert}},
  \bibinfo{author}{\bibfnamefont{R.}~\bibnamefont{Ochs}},
  \bibinfo{author}{\bibfnamefont{D.}~\bibnamefont{Beckmann}},
  \bibinfo{author}{\bibfnamefont{H.~B.} \bibnamefont{Weber}},
  \bibinfo{author}{\bibfnamefont{M.}~\bibnamefont{Mayor}}, \bibnamefont{and}
  \bibinfo{author}{\bibfnamefont{H.}~\bibnamefont{v.~L\"ohneysen}},
  \bibinfo{journal}{Phys.\ Rev.\ Lett.} \textbf{\bibinfo{volume}{88}},
  \bibinfo{pages}{176804} (\bibinfo{year}{2002}).

\bibitem[{\citenamefont{Zawadowski}(1980)}]{Zawadowski80}
\bibinfo{author}{\bibfnamefont{A.}~\bibnamefont{Zawadowski}},
  \bibinfo{journal}{Phys.\ Rev.\ Lett.} \textbf{\bibinfo{volume}{45}},
  \bibinfo{pages}{211} (\bibinfo{year}{1980}).

\bibitem[{\citenamefont{Abramowitz and Stegun}(1972)}]{Abramowitz}
\bibinfo{author}{\bibfnamefont{M.}~\bibnamefont{Abramowitz}} \bibnamefont{and}
  \bibinfo{author}{\bibfnamefont{I.~A.} \bibnamefont{Stegun}},
  \emph{\bibinfo{title}{Handbook of Mathematical Functions}}
  (\bibinfo{publisher}{Dover}, \bibinfo{address}{New York},
  \bibinfo{year}{1972}).

\bibitem[{\citenamefont{Anderson et~al.}(1970)\citenamefont{Anderson, Yuval,
  and Hamann}}]{Anderson70b}
\bibinfo{author}{\bibfnamefont{P.~W.} \bibnamefont{Anderson}},
  \bibinfo{author}{\bibfnamefont{G.}~\bibnamefont{Yuval}}, \bibnamefont{and}
  \bibinfo{author}{\bibfnamefont{D.~R.} \bibnamefont{Hamann}},
  \bibinfo{journal}{Phys.\ Rev.\ B} \textbf{\bibinfo{volume}{1}},
  \bibinfo{pages}{4464} (\bibinfo{year}{1970}).

\bibitem[{\citenamefont{Tsvelick and Wiegmann}(1983)}]{Tsvelick83}
\bibinfo{author}{\bibfnamefont{A.~M.} \bibnamefont{Tsvelick}} \bibnamefont{and}
  \bibinfo{author}{\bibfnamefont{P.~B.} \bibnamefont{Wiegmann}},
  \bibinfo{journal}{Adv. in Phys.} \textbf{\bibinfo{volume}{32}},
  \bibinfo{pages}{453} (\bibinfo{year}{1983}).

\bibitem[{\citenamefont{Sangregorio et~al.}(1997)\citenamefont{Sangregorio,
  Ohm, Paulsen, Sessoli, and Gatteschi}}]{Sangregorio97}
\bibinfo{author}{\bibfnamefont{C.}~\bibnamefont{Sangregorio}},
  \bibinfo{author}{\bibfnamefont{T.}~\bibnamefont{Ohm}},
  \bibinfo{author}{\bibfnamefont{C.}~\bibnamefont{Paulsen}},
  \bibinfo{author}{\bibfnamefont{R.}~\bibnamefont{Sessoli}}, \bibnamefont{and}
  \bibinfo{author}{\bibfnamefont{D.}~\bibnamefont{Gatteschi}},
  \bibinfo{journal}{Phys.\ Rev.\ Lett.} \textbf{\bibinfo{volume}{78}},
  \bibinfo{pages}{4645} (\bibinfo{year}{1997}).

\bibitem[{\citenamefont{Park and Pederson}(2004)}]{Park04add}
\bibinfo{author}{\bibfnamefont{K.}~\bibnamefont{Park}} \bibnamefont{and}
  \bibinfo{author}{\bibfnamefont{M.~R.} \bibnamefont{Pederson}},
  \bibinfo{journal}{Phys.\ Rev.\ B} \textbf{\bibinfo{volume}{70}},
  \bibinfo{pages}{054414} (\bibinfo{year}{2004}).

\end{thebibliography}

\end{document}